\documentstyle[psfig,12pt]{article}

\def\beq{\begin{eqnarray}}
\def\eeq{\end{eqnarray}}
\def\ln{\,\mbox{ln}\,}

\def\al{\alpha}
\def\be{\beta}
\def\ch{\chi}
\def\ga{\gamma}
\def\de{\delta}
\def\vp{\varepsilon}

\def\ka{\kappa}

\def\na{\nabla}

\def\si{\sigma}

\def\Ga{\Gamma}

\def\La{\Lambda}

\begin{document}

\centerline{\bf ON THE STATIC SOLUTIONS IN GRAVITY WITH}
\centerline{\bf MASSIVE SCALAR FIELD IN THREE DIMENSIONS}
\centerline{\bf }
\vspace*{0.37truein}
\centerline{G. de Berredo-Peixoto\footnotesize
\footnote{Present address: Departamento de F\'{\i}sica, Universidade Federal de 
Juiz de Fora, Juiz de Fora, MG, Brazil 36036-330. 
Electronic address: guilherme@fisica.ufjf.br } }
\centerline{\footnotesize\it Department of Physics, University
of Alberta, Edmonton, Canada.}
\centerline{\footnotesize\it 412 Avadh Bhatia Physics Laboratory,
Edmonton, AB, T6G 2J1 Canada.}

\vskip 10mm

{\centerline{\sl ABSTRACT}}
\vskip 3mm

\begin{quotation}
We investigate circularly symmetric static solutions in three-dimensional
gravity with a minimally coupled massive scalar field.
We integrate numerically the field equations assuming asymptotic
flatness, where black holes do not exist and a naked singularity is
present. We also give a brief review on the massless cases
with cosmological constant.
\end{quotation}



\vskip 8mm

\section{Introduction}

Black holes (BH) are one of the most intriguing features of General Relativity.
Although they are defined by geometrical concepts\footnote{see \cite{wald,frolov} for a
pedagogical introduction to General Relativity and black holes.}, they possess
remarkable thermodynamical properties, according to
the results of Hawking \cite{hawking75}. These properties arise due to quantum effects of
matter in a curved spacetime (Hawking radiation), providing strong identification between the
laws of thermodynamics and the classical energy formula in General Re\-la\-ti\-vi\-ty
(see \cite{bch73}),
which had been regarded only as a mathematical relationship.

In principle, the BH thermodynamics is not described by some statistical formalism, as
in the case of standard thermodynamics, but it is derived from geometrical degrees of
freedom. For example, the BH temperature is usually given by the surface gravity (which is a
geometrical object) for any reasonable matter content. However, it is not clear how the
BH termodynamics can be affected if one considers a gravity theory with additional degrees
of freedom, besides the metric. One can consider, for example, the Poincar\'e Gauge Theory
of Gravity (PGTG) \cite{hehl}, which includes
torsion in the physical description. For a review concerning theories with torsion,
see \cite{shapirotor}.

There are many possible approaches to this investigation. One could consider theories
with algebraic torsion (non-dynamical torsion). In this case, torsion is trivial
in the absence of spin fields. Thus, given a BH vacuum solution, all the consequent
thermodynamics is exactly the same as in General Relativity. On the other hand, one
can consider more interesting models, with dynamical torsion. One can find spherically
symmetric BH solutions with dynamical torsion in the four dimensional $R^2$-Gravity
framework, in the teleparallel limit \cite{baekler80} as well as in the more general cases
\cite{baekler81,hehl-aximmetric}. In Ref. \cite{hehl-aximmetric} one can find a stationary
BH solution in the quadratic Poincar\'e Gravity.

In this paper we consider three dimensional gravity coupled to scalar field, 
with special attention on black holes configurations. 
The first black hole solution in three dimensions was discovered by Ba\~nados,
Teitelboim and Zanelli \cite{btz}. Among many interesting properties of the BTZ black holes
(see the review \cite{carlip}), one can mention that it is a string solution in the low
energy regime \cite{horowitz}. In addition, the BTZ black hole is achieved as the final state
of the spacetime with collapsing matter \cite{mannross}, where one can find equilibrium
conditions for stable stars with matching conditions with the exterior BTZ solution
\cite{cruzzanelli}. See \cite{gil} for collapsing solution with scalar field without the
BTZ solution as the final state.  In the papers of Refs. \cite{btz2,clement} the generalization to
charged stationary solutions is investigated. One can find also the magnetic counterpart
of the BTZ static solution\cite{clement,welch,cataldo}, as well as its rotating
solution \cite{dias1}. The paper in Ref. \cite{dias2} extends the rotating electrically charged
BTZ solution to include Brans-Dicke theory. Solutions for minimally coupled massless scalar field
with non-trivial cosmological constant can be found in Ref. \cite{clement2}, 
without black holes. There are two classes of solutions: one with naked singularity and other
which is geodesically complete. In addition, one can find a one-parameter family of charged
black holes for the 3D-dilaton gravity \cite{chan-mann}, as well as its uncharged rotating
version \cite{chan-mann2}, where the black hole is specified by the mass, angular momentum and
the dilaton coupling parameters. A non-trivial black hole solution was found for a minimally coupling 
scalar field and a class of potentials \cite{gegenberg}. Recently, Garcia {\it et al} have developed 
a three-dimensional Poincar\'e Gauge Theory of Gravity including the Chern-Simons term \cite{hehlmacias}. 
Maluf and Sousa \cite{maluf} considered the 
restricted teleparallel version (with a Lagrangian quadratic in the torsion tensor) 
to describe a black hole solution following the Hamiltonian fomalism.
 
This paper is organized as follows. In Section 2, we introduce the general action for 
gravity minimally coupled to a massive scalar field, consider 
some special cases avaiable in the literature, and present the field equations
for a particular action, with vanishing cosmological constant. 
We show also a possible interpretation for the action in the Appendix. In Section 3, we find a solution by 
numerical integration, which has a naked singularity. We show that BH solutions
are absent (what is related to the no-hair theorems \cite{sudarsky}). 
Finally, in Section 4, we give our conclusions.

\section{Action and field equations}

Let us introduce the general action with a scalar field (minimally coupled 
to gravity) which has the following form:
\beq
S = \int\, d^3x\sqrt{-g}\left\{ \frac{1}{\kappa}R -
\na_\mu\phi\na^\mu\phi -M^2\phi^2 -\frac{1}{4}e^{a\phi}F_{\mu\nu}F^{\mu\nu} + V[\phi ]
\right\}\, , \label{generalaction}
\eeq
where $\ka$ is a coupling constant\footnote{Here, $\ka$ has dimensions of mass$^{(-1)}$, and,
in principle, it is not related to Newton's constant, due to the lack of Newtonian
limit in 3D-Einstein's theory \cite{giddings}.}.
This action describes a wide class of possible theories, and a complete study is
still not avaiable. Many authors have investigated some sectors of the action 
(\ref{generalaction}). For instance, Chan and Mann \cite{chan-mann} considered 
circularly symmetric static solutions in the massless case, $M = 0$, and they found
a class of charged black hole configurations, with $V[\phi ] = -2e^{b\phi}\La $. 
The rotating black holes were also discovered in the uncharged case 
\cite{chan-mann2}. If we let $b=0$, $M=0$ and $F_{\mu\nu}=0$, the resulting action 
is the one considered by Cl\'ement and Fabbri \cite{clement2}, and their solutions 
can not be achieved continuously from the dilaton solution \cite{chan-mann} by 
the procedure $b\to 0$.
We are going to investigate the uncharged version of (\ref{generalaction}), with 
$V[\phi ] =0$ and $M\neq 0$:
\beq
S = \int\, d^3x\sqrt{-g}\left\{ \frac{1}{\kappa}R -
\na_\mu\phi\na^\mu\phi -M^2\phi^2 \right\}\, .\label{action}
\eeq

Note that the massive case are not convenient for searching a black hole 
solution by the virtue of the non-scalar hair theorem \cite{bekenstein-prl}.
Surprisingly, no one has so far investigated the solutions for the theory
(\ref{action}), which is relatively simple in contrast to the more complicated 
cases already known in the literature. Let us remark that one should not
expect to cover other known solutions by taking $M\to 0$ (e.g. Ref. 
\cite{clement2} with $\La =0$).

One should mention that the motivations for studying theories with scalar
field varies from the Standard Model to Cosmology, as well as some aspects 
of the low energy regime of string theory. In the Appendix, we give a 
possible physical interpretation of the scalar field in the action 
(\ref{action}).       

\subsection{Field equations}

We are looking for circularly symmetric static solutions of theory (\ref{action}), 
thus there are two Killing fields,
$\partial /\partial t$ and $\partial /\partial\theta$ and the metric is given by
\beq
ds^2 = -f(r)\, dt^2 +\frac{g(r)}{f(r)}\, dr^2 + r^2\, d\theta ^2 \, .  \label{metric}
\eeq
The Killing field $\partial /\partial t$ is hypersurface orthogonal, because the
metric is static. Otherwise, the line element (\ref{metric}) would have crossing
terms of the type $h(r)\, dt\, d\theta$.

Varying the action (\ref{action}) with respect to the field variables 
($g_{\mu\nu}$, $\phi$), one can find the field equations
\beq
R_{\mu\nu} - \frac{1}{2}g_{\mu\nu}R = \ka T_{\mu\nu}
\eeq
and
\beq
\Box\phi - M^2\phi = 0 \, ,   \label{scalareq}
\eeq
where the momentum-energy tensor is given by
\beq
T_{\mu\nu} = \na_\mu\phi\na_\nu\phi -\frac{1}{2}\left( \, g_{\mu\nu}\na_\al\phi\na^\al\phi
+ M^2g_{\mu\nu}\phi ^2 \,\right) -\frac{1}{\ka}g_{\mu\nu}\La \, ,   \label{momentumenergy}
\eeq
and the operator $\Box$ stands for $g^{\al\be}\na_\al\na_\be$. Here we introduced, 
for the sake of generality, the cosmological constant through the procedure 
$e^{b\phi}\to 1/\ka$ in (\ref{generalaction}).  

After some calculations, one can achieve the following independent field equations
(assuming $\phi = \phi (r)$)
\beq
\frac{f^{'}}{2fr}-\frac{g^{'}}{2gr} = -\frac{1}{2}\ka (\phi^{'})^2 -
\frac{1}{2}\ka\frac{g}{f}M^2\phi ^2 - \frac{g}{f}\La\, ;       \label{e1}
\eeq
\beq
-\frac{f^{'}}{2fr} = -\frac{1}{2}\ka (\phi^{'})^2 +
\frac{1}{2}\ka\frac{g}{f}M^2\phi ^2 + \frac{g}{f}\La\, ;       \label{e2}
\eeq
\beq
\phi ^{''} + \left(\, \frac{f^{'}}{f}-\frac{g^{'}}{2g}+\frac{1}{r}\,\right) \,\phi ^{'} +
\frac{g}{f}M^2\phi = 0 \, ,                                 \label{e3}
\eeq
where $f^{'}$ means $df/dr$. After some manipulations, we obtain the equivalent set of field equations
\beq
\frac{g^{'}}{2gr} = \ka (\phi^{'})^2\, ; \label{e11}
\eeq
\beq
-\frac{f^{'}}{2fr} = -\frac{1}{2}\ka (\phi^{'})^2 + \frac{1}{2}\ka\frac{g}{f}M^2\phi ^2
+ \frac{g}{f}\La \, ;   \label{e22}
\eeq
\beq
\phi ^{''} - \ka\frac{g}{f}M^2\phi ^2\phi ^{'}r - 2\frac{g}{f}\La\phi ^{'}r +
\frac{1}{r}\phi ^{'} + \frac{g}{f}M^2\phi = 0\, . \label{e33}
\eeq
Notice that by (\ref{e11}) and (\ref{e33}), solutions of the type $g(r) = {\rm const.}$
are possible only if the field $\phi (r)$ is trivial, producing thus the well known
static BTZ black hole solution \cite{btz} (See \cite{carlip} for a review on the classical
and quantum properties of the BTZ black hole).

\section{Numerical integration}

Before considering general solutions, let us investigate some particular cases
to achieve more insight in the search for the general solution. In the case $M = \La =0$,
equation (\ref{e33}) reads
\beq
\phi ^{''} + \frac{1}{r}\phi ^{'} = 0 \, ,
\eeq
which has the solution
\beq
\phi (r) = Q\,\ln\frac{r}{r_0}\, ,  \label{phi}
\eeq
where $Q$ and $r_0$ are integration constants, with canonical dimensions $+1/2$ and $-1$
respectively. Other equations can be integrated, resulting in the following solution:
\beq
g(r) = \ga r^{2\al}\, ,\;\;\; {\rm with}\;\; \al := \ka Q^2\, {\rm and}  \label{g}
\eeq
\beq
f(r) = \de r^{\al}\, ,
\eeq
where $\ga$ and $\de$ are integration constants with canonical dimensions $2\al$ and $\al$. In
this solution, the diverging scalar field provides diverging metric
functions $f(r)$ and $g(r)$,
far away from the origin, $r>>1$, with no horizon formation. In the flat background, $\phi$ has the
same solution (\ref{phi}). Now consider
the Minkowski spacetime with $M\neq 0$. In four dimensions, the solution is the well known meson
field, $\phi = Q{\rm exp}(-Mr)$. In our case, the equation for $\phi$ is a Bessel equation,
with the modified Bessel function of the second kind as a suitable solution,
\beq
\phi (r) = Q\, K_0(Mr)\, ,
\eeq
with similar asymptotic behaviour to the meson solution in four dimensions.

We cannot find analytical solutions for the general case with $\La =0$. Thus we integrate
numerically using the software MAPLE\footnote{The reader can obtain useful information
about the numerical integration applied to General Relativity in several papers, say, for example, in papers of Ref. \cite{proca-einstein}.} 
(See, for instance, the book \cite{maple}). 
We first guess suitable initial values for the
fields at infinity and then integrate from that point to the origin. As the mass is non-trivial, one
can suppose that the scalar field decays with $r$. At infinity, we assume that
the metric is approximately free from the influences of the scalar field, thus the metric at infinity
can be regarded as the vacuum solution of the Einstein's equations. The appropriate
ansatz for the metric at infinity is (see \cite{giddings})
\beq
ds^2 = -dt^2 + b \, dr^2 +r^2d\theta ^2
\eeq
which describes a flat spacetime with conical singularity at the origin ($b=$ constant). To
make this supposition, we assume that the solution is asymptotically flat.
The field equation
for $\phi (r)$ in this background has solution $\phi (r) = Q\, K_0(M\sqrt{b}r)$. Let 
us remark that this ansatz is valid only in the asymptotic region; thus, despite this
function is singular at the origin, this feature may not be true for the actual solution
satisfying $\phi (r) \sim K_0(M\sqrt{b}r)$ far from the origin. We choose
$r=30$ as the infinity and set $Q=2$, $M=0.1$, $b=1.1$, $f(30)=1$, $g(30)= b$,
$\phi (30) =Q\, K_0(30M\sqrt{b})$ and $\phi^{'}(30)=d\phi /dr|_{r=30}$. The
numerical integration carried out is described by the plotting on Figure 1. As we increase
the parameter $M$, the deviations of functions $f(r)$ and $g(r)$ from the vacuum solution,
near origin, are decreased even more. These deviations are continuously increased if the
scalar "charge", $Q$, is increased (as far as the value $r=30$ can be regarded as infinity
with a negligible error). There is no black hole, as shown by Figure 1.

One can also integrate from the origin, starting from a suitable initial conditions.
To find them, one has to guess an approximate solution near $r=0$. If we assume
regularity in the origin, we suppose a power series in $r$ for the variables:
\beq
f\approx f_0+f_1r+f_2r^2+...\, ;\;\;\;\; g\approx g_0+g_1r+g_2r^2+...\; ; \nonumber
\eeq
\beq
\phi\approx \phi_0+\phi_1r+\phi_2r^2+... \label{regularorig}
\eeq
Notice that this guess has a limited importance, specially if some black hole
solution is supposed to be found. However, it is a possible procedure of 
numeric integration from the origin. 


One can verify that these solutions satisfy the field equations up to $O(r^2)$,
if some constraints between the constant parameters are taken into account. After that,
we find the following approximate solution for small values of $r$:
\beq
f(r)=a-\frac{1}{2}bq^2r^2\, ;\;\;\;  g(r)=b\, ;\;\;\;  
\phi (r)=q-\frac{1}{4}\frac{bq}{a}r^2\, ,
\eeq
where $a$, $b$ and $q$ are constant parameters. Thus, there is a three-parameter family of
possible solutions near the origin. As the asymptotic solution at infinity has
two parameters, only a sub-family of the tree-parameter family would produce (in
principle) a convergent solution which coincides with the solution at infinity.
However, we shall show that this three-parameter family is incompatible with the
asymptotic flatness, so both families are actually distinct. Now,
let us consider an approximate solution near an hypothetic horizon. One can write
the following ansatz (assuming regularity of all functions at the horizon, $r_h$)
\beq
f & \approx & f_0(r-r_h)+f_1(r-r_h)^2+...\; ; \nonumber \\
g & \approx & g_0+g_1(r-r_h)+g_2(r-r_h)^2+...\; ;  \label{regularhoriz} \\
\phi & \approx & \phi_0+\phi_1(r-r_h)+\phi_2(r-r_h)^2+...   \nonumber
\eeq
Again, one substitute these expressions into the field equations and demand that
they must be satisfied up to the second order in $(r-r_h)$. We can achieve the
following solution:
\beq
f= a(r-r_h)^2\, ;\;\;\; g=b+c(r-r_h)^2\, ;\;\;\;  \phi = \phi^{'}=0\, .
\eeq
Actually this solution corresponds to the trivial scalar field case, and should not be
considered. Now we shall
show that no black holes are possible if the spacetime is asymptotically
flat\footnote{The author is grateful to Professor Don Page for explaining this point.},
unless the scalar field is trivial.

\subsection{No massive scalar hair}

Consider the static circularly symmetric line element, in some appropriate coordinate system,
\beq
ds^2 = -U^2(x)dt^2+dx^2+R^2(x)d\theta ^2\, .
\eeq
The minimally coupled scalar field, $\phi (x)$, is a solution of (\ref{scalareq}):
\beq
\frac{1}{UR}(UR\phi^{'})^{'}=M^2\phi \, ,
\eeq
where the prime means derivative with respect to $x$.
Let us assume regularity of $\phi$ at the horizon ($U=0$) or at the origin ($R=0$).
Thus, we must have $\phi^{'}=0$ there. Without loss of generality, one can make
$\phi >0$ there, but not $\phi=0$ (in this case, $\phi$ is zero everywhere). Then
the quantity $(UR\phi^{'})^{'}$ is positive as we move outside the horizon or the origin.
One can write
\beq
UR\phi^{'}|^{x}_{x_0}=\int_{x_0}^xdx(M^2UR\phi )  \, .
\eeq
where the lower limit $x_0$ represents the horizon or the origin. As $\phi^{'}$ is positive
(according to the previous equation), we obtain
\beq
UR\phi^{'}|^{x}_{x_0}>\phi (x_0)\int_{x_0}^xdx(M^2UR)\, . \label{ineq}
\eeq
If the spacetime is asymptotically flat, then $U\to 1$ and $R\to x$ as $x\to\infty$.
So we can consider the dominant contribution to (\ref{ineq}) for large $x$, as follows:
\beq
x\phi^{'}>\frac{1}{2}M^2x^2\phi (x_0)   \, .
\eeq
Consequently, we have $\phi^{'}>(1/2)M^2 x\phi (x_0)$, which means that the scalar field grows too
fast, so it will prevent the spacetime from being asymptotically flat by the back reaction.
Thus, there is no static circularly symmetric asymptotically flat solution for the
non-trivial massive scalar field minimally coupled with Einstein gravity (if the scalar field
is regular everywhere). Also, the same
result is valid for the asymptotically De Sitter and anti-De Sitter cases. The asymptotically
anti-De Sitter spacetime is particularly interesting because it can be regarded as some
asymptotic solution (e.g., static BTZ solution) for a black hole solution with
massive scalar field and $\La < 0$.

This result implies that both the ansatz (\ref{regularorig}) and (\ref{regularhoriz})
are incompatible with the solution assumed at the infinity, which is asymptotically flat,
because they are approximate solutions with regular scalar field at the origin and at the
horizon, respectively. Thus the integration carried out from infinity should
correspond to a solution with non-regular scalar field at the origin. In this case,
there is a naked singularity.

\section{Conclusions}

We have investigated circularly symmetric static solutions of gravity minimally 
coupled to massive scalar field in three dimensions, and then we 
presented a numerical solution containing a naked
singularity and consistent with the no-hair theorems and no-black hole
theorem (see \cite{ida}) in three dimensions. 

By the results of subsection \S 3.1, the unique possibility to find a non-trivial 
black hole solution is to consider the cases with non-regular scalar field at 
the origin. Among these cases, there are two classes of solutions: \\

(i) Asymptotically flat solutions; \\

(ii) Solutions which are not asymptotically flat. \\

The first group are covered by the numerical integration procedure 
described in the preceding section. The second group (ii) was not
considered in this paper. Notice that in principle one can not guess
suitable initial conditions in any region.

\vskip 5mm
 \noindent {\bf Acknowledgments.} \vskip 3mm
I would like to thank Professor V.P. Frolov for invaluable
discussions and to University of Alberta for warm hospitality and support. I also
acknowledge Professor J.A. Helay\"el-Neto and Professor D. Page for useful discussions.
This work was done with support from CNPq, a Brazilian
Government institution that promotes development in science and technology.


\begin{appendix}
\section*{Appendix}
\subsection*{Gravity with dynamical torsion in three dimensions}

Let us consider Dirac fermions in the Riemann-Cartan three-dimensional spacetime,
which can be described by the action ($\hbar = c = 1$)
\beq
S_{\psi} = \int d^3x\sqrt{-g}\left\{ \frac{1}{\ka}\tilde{R} +
\frac{i}{2}(\bar{\psi}\ga^\mu\tilde{\na}_\mu
\psi - \tilde{\na}_\mu\bar{\psi}\ga^\mu\psi )-m\bar{\psi}\psi \right\} \, ,
\eeq
where the objects with tilde are constructed with non-trivial torsion
($R$ is the Ricci scalar and $\na_\mu $ is the Riemannian covariant derivative),
$$
T^\mu\mbox{}_{\al\be} := \tilde{\Ga}^\mu\mbox{}_{\al\be} - \tilde{\Ga}^\mu\mbox{}_{\be\al}\, .
$$
We adopt $R^\mu\mbox{}_{\nu\al\be}:=\partial_\al\Ga^\mu\mbox{}_{\nu\be} -\, ...$.
The covariant gamma matrices are defined in the usual way,
$\ga^\mu := e^\mu\mbox{}_a\,\ga^a$ ($e^\mu\mbox{}_a$ are the dreibeins).
By the procedure $\na_\mu\to\tilde{\na}_\mu $, we
consider minimal coupling between torsion and fermions.
Straightforward calculations enable us to write\footnote{See \cite{gasperini} for detailed
calculations, and \cite{helayel} for the Minkowski background case in three dimensions.}
\beq
S_{\psi} = \int d^3x\sqrt{-g}\left\{ \frac{1}{\ka}\tilde{R}+
\frac{i}{2}(\bar{\psi}\ga^\mu\na_\mu
\psi - \na_\mu\bar{\psi}\ga^\mu\psi )-m\bar{\psi}\psi 
+  \frac{i}{8}T_{\mu\nu\al}\bar{\psi}\ga^{[\mu}\ga^\nu\ga^{\al ]}\psi\right\}\, .
\eeq
Here the brackets mean total antisymmetrization.
We can choose the following gamma matrices basis, $\ga^a$,  in terms of the Pauli matrices, $\si^i$,
\beq 
\ga^0:=-i\si^3\, ; \;\;\; \ga^1:=\si^1\, ; \;\;\; \ga^2:=\si^2\, .
\eeq
They satisfy
\beq
\left\{ \,\ga^\mu\, ,\, \ga^\nu\,\right\} = 2g^{\mu\nu}\,\hat{1}\,
\eeq
where $g^{\mu\nu}$ has signature $+1$. After some algebra, one can achieve
\beq
\ga^{[\mu}\ga^\nu\ga^{\al ]} = \frac{1}{\sqrt{-g}}\ga^0\ga^1\ga^2\vp^{\mu\nu\al} =
\frac{1}{\sqrt{-g}}\vp^{\mu\nu\al}\, ,
\eeq
where $\vp^{\mu\nu\al}$ is the antisymmetrization symbol ($\vp^{012}:=+1$).
Thus, taking into account that the totally antisymmetric part of the torsion in three
dimensions is related to the pseudo-scalar $\ch := \vp^{\mu\nu\al}T_{\mu\nu\al}$
(which has weight $-1$), we can write the interacting action as
\beq
S_I=\int d^3x\sqrt{-g}\frac{i}{8}T_{\mu\nu\al}\bar{\psi}\ga^{[\mu}\ga^\nu\ga^{\al ]}\psi =
\int d^3x\sqrt{-g} \frac{i}{8}\bar{\psi}\frac{1}{\sqrt{-g}}\ch\psi \, ,
\eeq
where $\frac{1}{\sqrt{-g}}\ch$ is a scalar.
Thus, in three dimensions, the minimal coupling between fermions and torsion gives a Yukawa
coupling, and the torsion component which couples with fermions is described by the
pseudo-scalar $\ch (x)$ \cite{helayel}.

Now we consider only the $\ch$-component of torsion in the description of gravity with
dynamical torsion. We choose the action
\beq
S = \int d^3x\sqrt{-g}\left\{ \,\frac{1}{\ka}R - \na_\mu\ch\na^\mu\tilde{\ch} -
M^2\ch\tilde{\ch}\,\right\}\, ,
\eeq
where $\tilde{\ch}$ is the dual of $\ch$, $\tilde{\ch} := -\frac{1}{g}\ch$, such that the action
is a scalar. After the variable change
\beq
\ch (x) \to \phi (x)= \frac{1}{\sqrt{-g}}\ch (x)\, ,
\eeq
the action can be written as a scalar-metric action (\ref{action}). Notice that the conventional
theories with dynamical torsion are usually described by an action in a different way. 
For instance, one considers the Lagrangian quadratic in torsion and total curvature,
which are the field strength of tetrads and spin connection, respectively (see \cite{hehl}).
After some manipulation, it is possible to write the usual Lagrangian in terms of the metric
and the torsion, such that the Lagrangian is written as a torsion component plus a purely 
Riemannian one. In the theory (\ref{action}) the dynamical term for the torsion was 
introduced by hand, and not in the natural way from to the total curvature squared. 
For theories of dynamical torsion in this scenario, see, for example, the papers in the
Refs. \cite{jhep,shapirotor}.

\end{appendix}

\begin{center}
***
\end{center}

Figure caption: Numerical integration from infinity, 
assuming the asymptotically flat ansatz. The curves $G$, $F$ 
and $U$ correspond to the functions $g(r)$, $f(r)$ and $\phi (r)$, 
respectively.

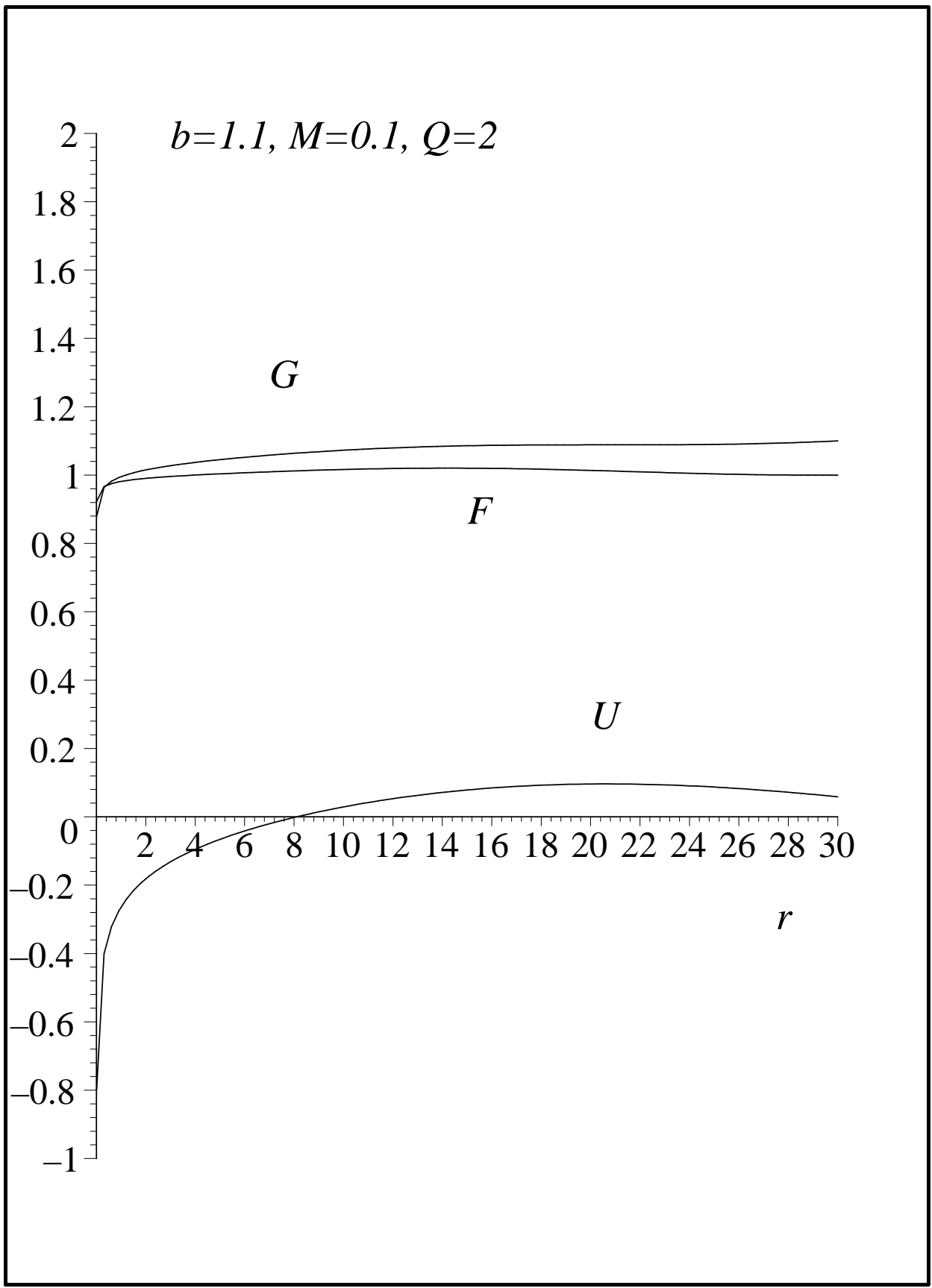
\end{document}